\documentclass{blois}

\def\be{\begin{equation}}
\def\ee{\end{equation}}
\def\bea{\begin{eqnarray}}
\def\eea{\end{eqnarray}}

\newcommand{\Photo}{\includegraphics[height=35mm]{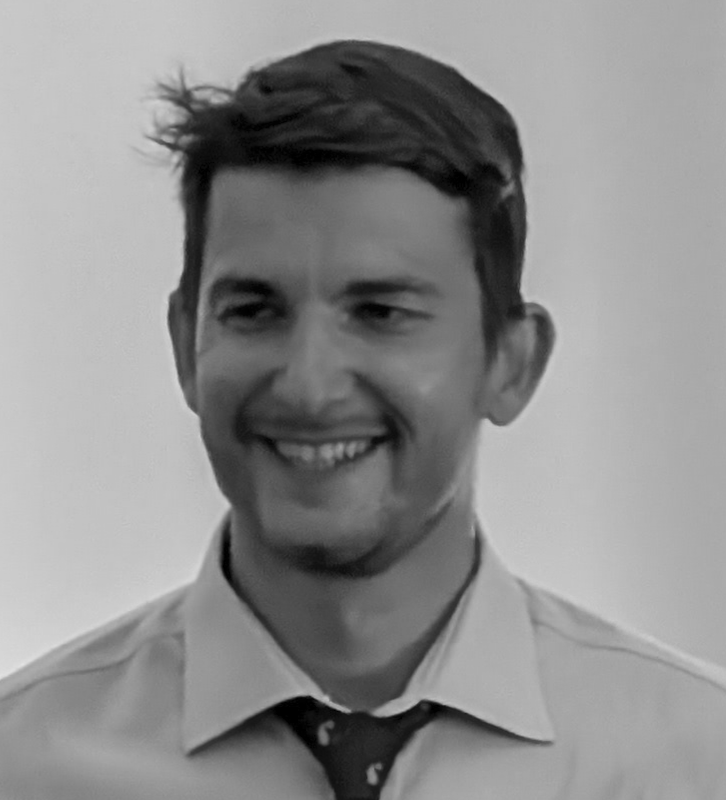}}

\begin{document}
\vspace*{4cm}
\title{Combined KM3NeT/ARCA and ANTARES search for compact neutrino sources}

\author{V.~Kulikovskiy$^{a}$, B.~Caiffi, S.~Alves Garre, J.~Aublin, M. de Jong, A.~Heijboer, G.~Illuminati, R.~Muller, V.~Parisi, M.~Sanguineti, F. Vazquez de Sola and S.~Zavatarelli \\ on behalf of the ANTARES and KM3NeT collaborations}

\address{$^{a}$INFN, Sezione di Genova, \\
Via Dodecaneso 33, Genova, 16146 Italy}

\maketitle\abstracts{
High-energy neutrinos from point-like sources are searched in the combined dataset from ANTARES and the available KM3NeT/ARCA data. This analysis demonstrates the potential of combining data from these neutrino  detectors using a binned likelihood approach.}

\section{Introduction}
KM3NeT/ARCA, under construction off Sicily, will comprise two building blocks of about 100 detection lines each, with 18 digital optical modules per line. Each module hosts 31 3-inch PMTs, that enhances directional resolution. Its predecessor, ANTARES, operated from 2008 to 2022 off Toulon, France, equipped with 12 lines, each hosting 25 storeys of three 10-inch PMTs. Both detectors offer optimal visibility of the Southern Sky and Galactic Centre.

In this joint analysis four datasets are used: two complete ANTARES datasets with a detector livetime of 4541 days~\cite{ANTARESdataset} split into non-overlapping track and shower datasets; ARCA19 tracks: track-like events from the KM3NeT/ARCA detector with 19 detection lines  (48.4 days)~\cite{ARCA19and21dataset}; ARCA21 tracks: track-like events from the KM3NeT/ARCA detector with 21 detection lines  (287.4 days)~\cite{ARCA19and21dataset}. Tracks are events where the production of a muon is involved.  Showers, instead, are mainly induced by the neutral-current interactions of neutrinos of all flavours, and by the charged-current interactions of electron neutrinos.

\section{Signal and background models and likelihood formalism}

The histogram of the observed events, $N$, is compared versus a scalable estimate of the number of signal events, $S$,  expected for a reference flux, $\Phi_0$, and the number of background events, $B$, as follows:
 \begin{equation}
    \ln L = \sum_i^{\rm N_{bins}} N_i \ln ( B_i +\zeta S_i ) -(B_i + \zeta S_i),
\end{equation}
where $\zeta$ is the signal strength, namely a free normalisation parameter for the signal with respect to the reference flux, $\Phi_0$. The chosen reference flux of one flavour $\nu+\bar\nu$ is $\Phi_0=10^{-8}$ GeV\,cm$^{-2}$\,s$^{-1}$ and a 1:1:1 flavour ratio is assumed together with equal neutrino-antineutrino composition.

The logarithm of the likelihood ratio between signal and non-signal hypothesis is used as a test statistic for $p$-value calculation: 
\begin{equation}
    \lambda = \ln L( \zeta=\hat{\zeta} ) - \ln L ( \zeta = 0 ),
\label{eq:lambda}
\end{equation}
where $\hat{\zeta}$ is the fitted signal strength and $\zeta\ge0$. To compute upper limits the following test statistic as described by Eq. (16) in~\cite{Cowan2011} is used:
\begin{equation}
\lambda_\mathrm{UL}=\left\{
\begin{array}{rl}
\ln{\frac{L(\zeta)}{L(0)}}, & \hat{\zeta}<0 \\
\ln{\frac{L(\zeta)}{L(\hat{\zeta})}}, & 0\le\hat{\zeta}\le\zeta \\
0, & \hat{\zeta}>\zeta
\end{array}
\right.
\end{equation}

The $5\sigma$ discovery potential is defined as the signal strength value~$(\zeta_{5\sigma})$ needed to obtain a p-value less than an equivalent $5\sigma$ excess over the background-only hypothesis ($\zeta=0$) in $50\%$ of the pseudo-experiment realisations with $\zeta=\zeta_{5\sigma}$. A one-sided Gaussian convention is used for p-value conversion. Discovery flux is calculated as $\Phi_{5\sigma} = \zeta_{5\sigma} \Phi_0$. The sensitivity is defined as the median upper limit on $\zeta$ at the 90\% confidence level~(CL) calculated with the Neyman approach using $\lambda_\mathrm{UL}$. It is converted to the $\Phi_{90}$ in the same way as as for the $5\sigma$ discovery flux.

The test statistics can be calculated summing over all bins of all four datasets in order to make a combined analysis. The $\lambda$ distributions are built for each true signal strength $\zeta$. In this analysis we distinguish ``generation'' histograms $S$ and $B$ used to create pseudo-data sets with Poisson sampling from ``evaluation'' histograms $S$ and $B$ used to calculate the test statistics $\lambda$.

Two-dimensional histograms as a function of the logarithm of the angular distance to the source centre, $\log_{10}{\alpha}$, and the logarithm of the reconstructed energy, $\log_{10}{E}$, are used for ARCA track and ANTARES shower events. Three-dimensional histograms as a function of $\log_{10}{\alpha}$, $\log_{10}{E}$, and the logarithm of the angular error estimation, $\log_{10}\beta$, are used for the ANTARES
track events. 

For all datasets the signal generation and evaluation histograms, $S$, are computed for each source declination using simulation. The background histograms, $B$, are obtained by populating data or simulated events from the same declination band. The band widths are optimised to ensure sufficient statistics. Empty bins are filled with the minimum non-zero value within the corresponding $\log_{10}E$ slice to ensure numerical stability during likelihood evaluation. Alternatively, a smoothed extrapolation using the \texttt{SUFTware} package is applied.

In total, four independent analyses were performed:
\begin{itemize}
\item \textbf{Case 1:} For the ARCA sample, the background estimate is entirely derived from data for both generation and evaluation. For the ANTARES samples, the background used for generation is data-driven, while the evaluation relies on Monte Carlo simulations.
\item \textbf{Case 2:} As in Case 1, with the difference that, for the ANTARES samples, the data used in the generation of the background are randomised, assuming a uniform distribution in the azimuthal angle and recalculating the declination.
\item \textbf{Case 3:} For all datasets, the background estimation is obtained using the DEFT algorithm, and consistently applied to both generation and evaluation.
\item \textbf{Case 4:} For the ARCA sample, the background estimate is data-driven for both generation and evaluation, while for the ANTARES samples it is entirely based on simulations.
\end{itemize}
The analysis is performed across declination bands and candidate sources under these four assumptions. The spread of the results is used to define a systematic uncertainty band associated with the background modelling.

\section{Results} 
\begin{figure}[htbp]
    \centering \includegraphics[width=0.49\linewidth]{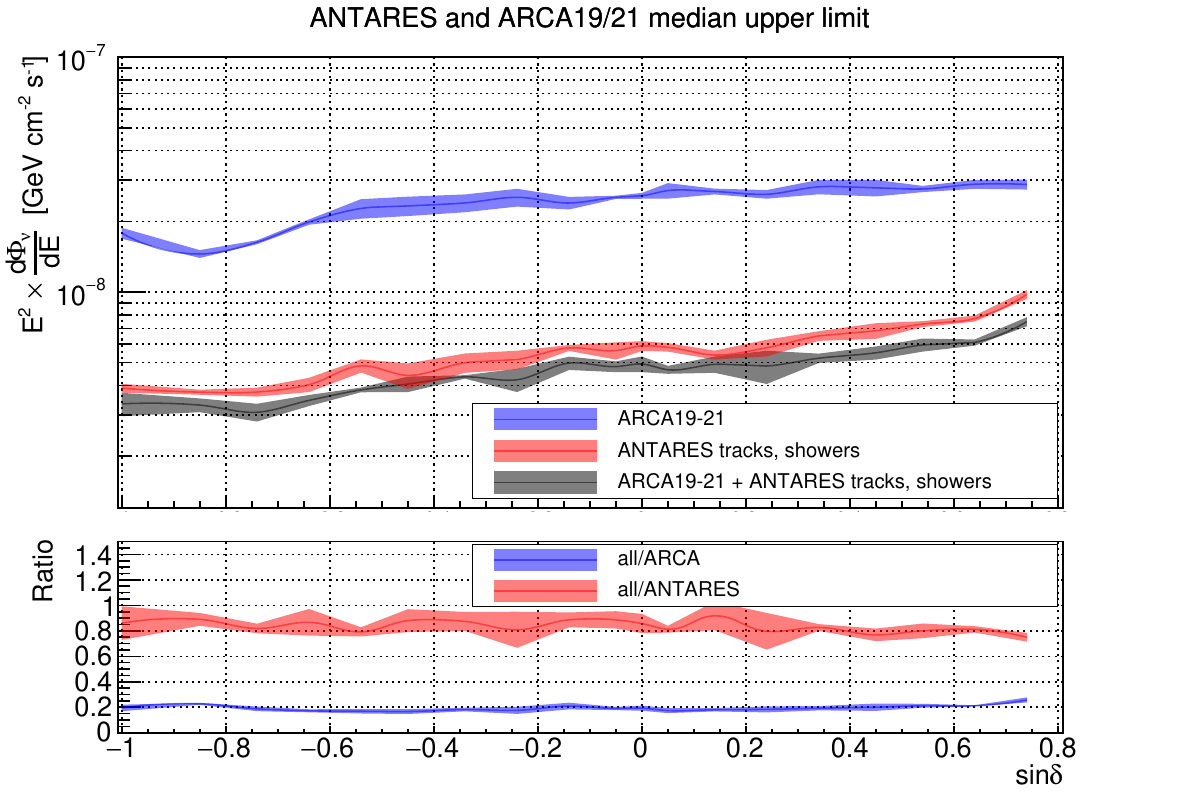}
    \includegraphics[width=0.49\linewidth]{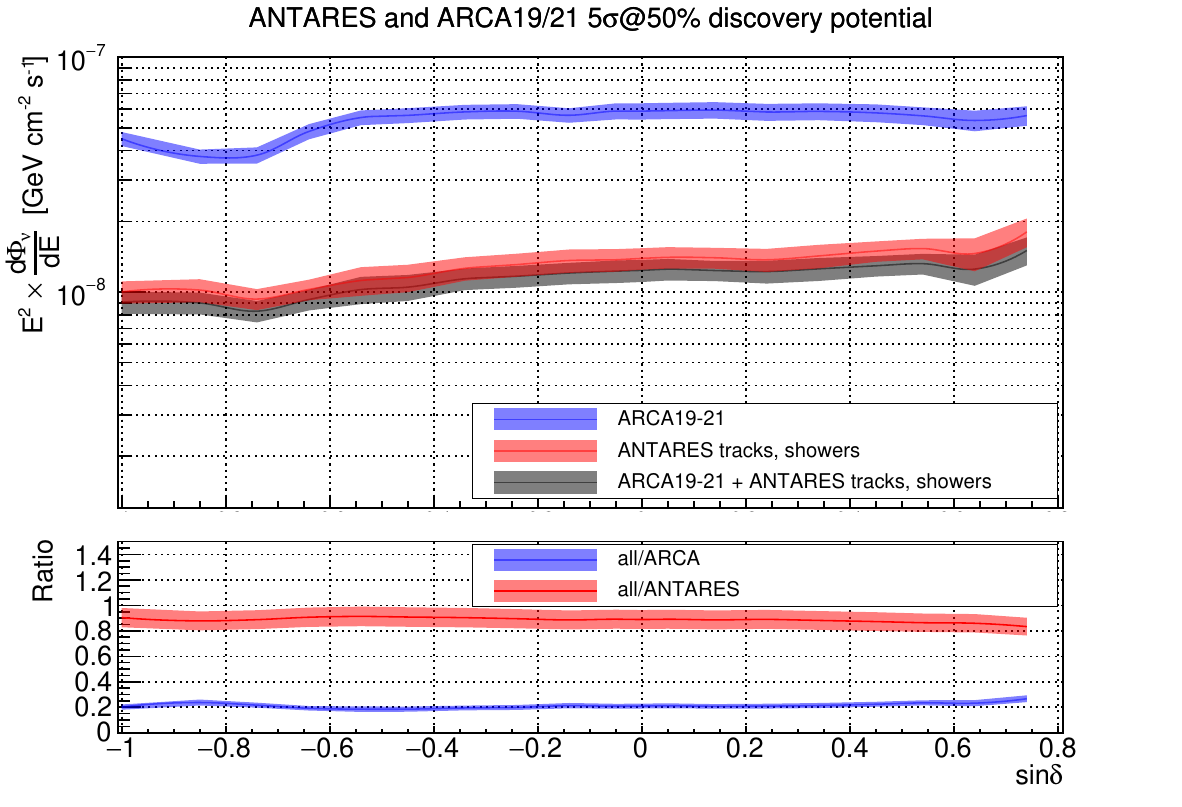}
    \caption{Left: 90\% CL median upper limits on the one-flavour neutrino flux as a function of declination. \\ 
    Right: 50\% 5$\sigma$ discovery flux on the one-flavour neutrino flux as a function of declination.}
    \label{fig:SensiDisco}
\end{figure}

The point-source sensitivity and discovery fluxes are shown in Fig.~\ref{fig:SensiDisco} for ARCA19-21, ANTARES, and the joint analysis assuming an $E^{-2}$ energy spectrum. In the current work, ANTARES significantly drives the results in the joint analysis. Nevertheless, combining with ARCA the performance improves by about 10\% with respect to the ANTARES standalone analysis. 

\begin{figure}
    \centering
    \begin{minipage}{0.49\textwidth}

      \centering
      \includegraphics[width=\linewidth]{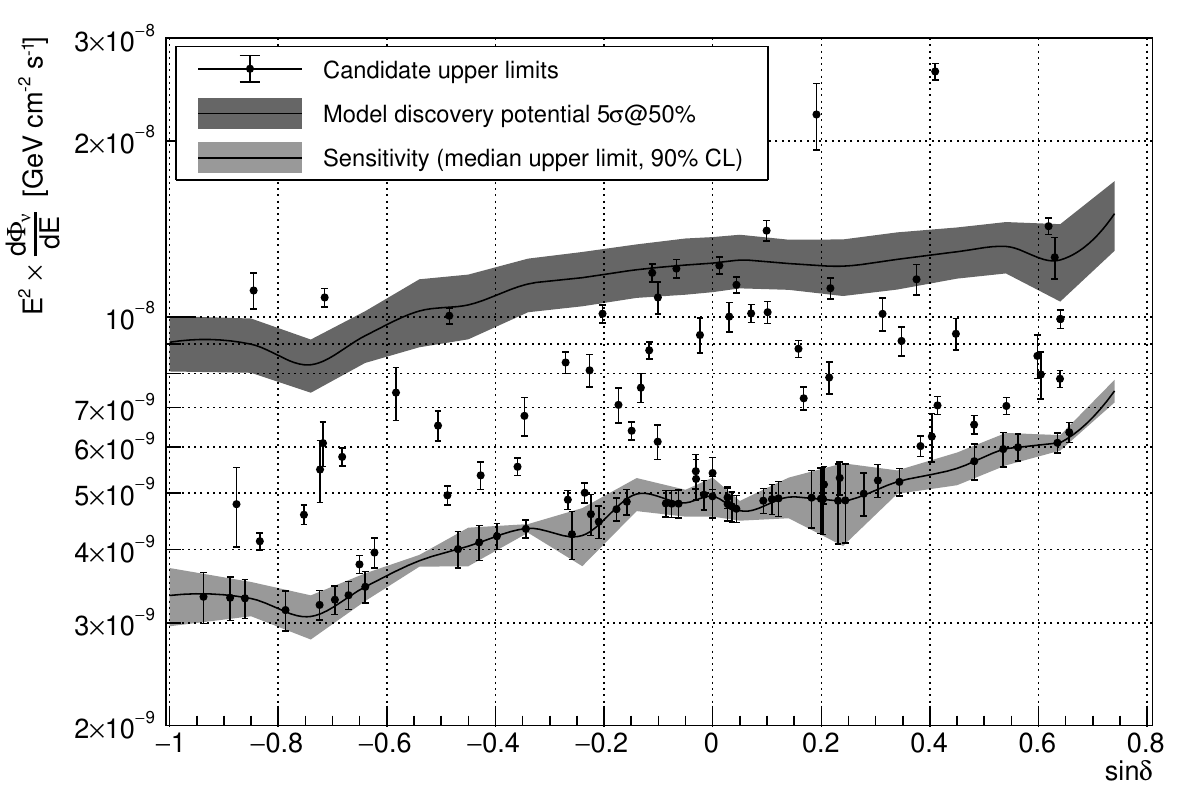}
     \caption{90\% CL upper limits (dots) on the one-flavour neutrino flux normalisation for the investigated astrophysical candidates as a function of the source declination, assuming an $E^{-2.0}$ spectrum.
}
    \label{fig:LimitsCandidates}
    \end{minipage}\hfill
    \begin{minipage}{0.49\textwidth}
    \centering
    \includegraphics[width=\linewidth]{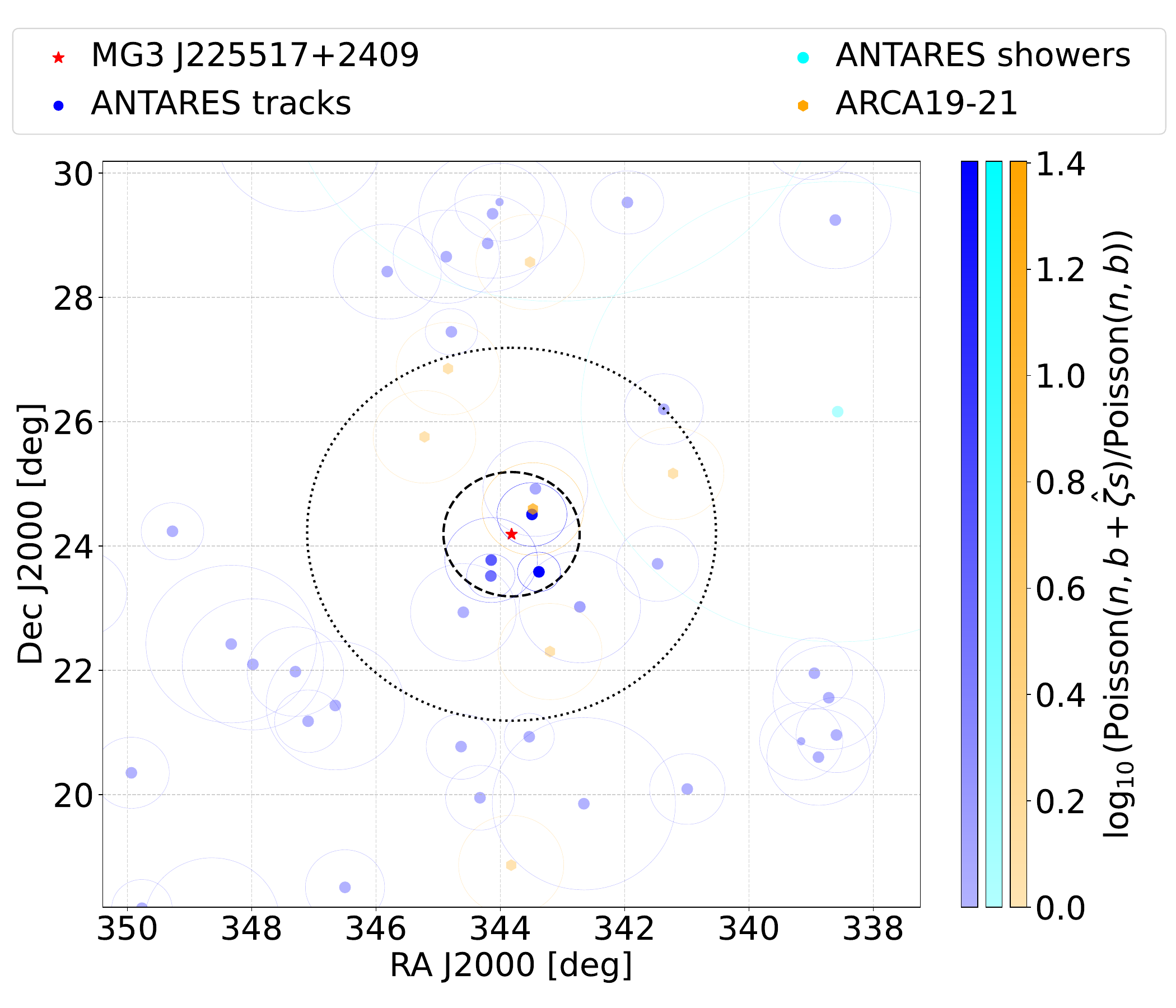}
    \caption{MG3~J225517+2409 sky map. The marker opacity corresponds to a test statistic for one bin, the circles around markers represent event angular uncertainty. The dashed black circles show 1$^\circ$ and $3^\circ$.
    }
    \label{fig:Maps}
    \end{minipage}
\end{figure}

The candidate-source list used in this analysis is inherited from the KM3NeT/ARCA analysis~\cite{ARCA19and21dataset}. The 90\% CL upper limits for each source are shown in Fig.~\ref{fig:LimitsCandidates} and in Table~\ref{t:results}.

\begin{table}
\centering
\resizebox{\textwidth}{!}{

\begin{tabular}{|lcccccc|lcccccc|}
\hline
Name & $RA$ [\textdegree] & $\delta$ [\textdegree] & Ext [\textdegree] & $\hat{\zeta}$ & p-value  & $\phi^{90\%}_\mathrm{CL}$ & Name & $RA$ [\textdegree] & $\delta$ [\textdegree] & Ext [\textdegree]& $\hat{\zeta}$ & p-value  & $\phi^{90\%}_\mathrm{CL}$ \\
\hline
LMC N132D & 81.2 & --69.65 & - & - & - & $0.33 \pm 0.03$ & J0725-0054 & 111.46 & --0.92 & - & - & - & $0.50 \pm 0.03$ \\
Danks 1 & 198.12 & --62.69 & 0.58 & - & - & $0.33 \pm 0.03$ & 1LHAASO J1848-0001u & 282.19 & --0.02 & - & - & - & $0.49 \pm 0.04$ \\
NGC 3603 & 168.79 & --61.26 & - & $0.02 \pm 0.02$ & $0.6 \pm 0.4$ & $0.48 \pm 0.07$ & NGC 1068 & 40.67 & --0.01 & 0.12 & - & - & $0.54 \pm 0.03$ \\
SNR G318.2+00.1 & 224.42 & --59.46 & 0.31 & - & - & $0.33 \pm 0.02$ & J2136+0041 & 324.16 & 0.7 & - & $0.371 \pm 0.007$ & $0.057 \pm 0.007$ & $1.23 \pm 0.02$ \\
Westerlund 2 & 155.99 & --57.7 & 0.75 & $0.33 \pm 0.07$ & $0.10 \pm 0.04$ & $1.11 \pm 0.07$ & J1058+0133 & 164.62 & 1.57 & - & - & - & $0.49 \pm 0.01$ \\
IC-hotspot South hemisphere & 350.2 & --56.5 & - & $0.0002 \pm 0.0002$ & $0.6 \pm 0.4$ & $0.414 \pm 0.006$ & J0108+0135 & 17.16 & 1.58 & - & - & - & $0.48 \pm 0.03$ \\
HESS J1614-518 & 243.54 & --51.87 & - & - & - & $0.32 \pm 0.02$ & PKS 0215+015 & 34.45 & 1.75 & - & $0.12 \pm 0.04$ & $0.14 \pm 0.01$ & $1.00 \pm 0.05$ \\
PKS 2005-489 & 302.37 & --48.82 & - & - & - & $0.46 \pm 0.01$ & J1229+0203 & 187.28 & 2.05 & - & - & - & $0.47 \pm 0.02$ \\
RX J0852.0-4622 & 133 & --46.37 & 0.63 & - & - & $0.32 \pm 0.02$ & TXS 0310+022 & 48.3 & 2.5 & - & - & - & $0.47 \pm 0.02$ \\
HESS J1641-463 & 250.26 & --46.3 & - & $0.03 \pm 0.03$ & $0.6 \pm 0.4$ & $0.55 \pm 0.07$ & 3C403 & 298.07 & 2.51 & - & $0.454 \pm 0.004$ & $0.0015 \pm 0.0004$ & $1.13 \pm 0.01$ \\
Westerlund 1 & 251.76 & --45.85 & 0.61 & $0.06 \pm 0.04$ & $0.21 \pm 0.05$ & $0.61 \pm 0.05$ & CGCG 420-015 & 73.35 & 4.06 & - & $0.301 \pm 0.006$ & $0.023 \pm 0.004$ & $1.01 \pm 0.02$ \\
VelaX & 128.75 & --45.6 & 0.58 & $0.45 \pm 0.02$ & $0.012 \pm 0.003$ & $1.08 \pm 0.02$ & J0433+0521 & 68.3 & 5.35 & - & - & - & $0.49 \pm 0.02$ \\
PKS 0537-441 & 84.71 & --44.08 & - & - & - & $0.33 \pm 0.02$ & TXS 0506+056 & 77.35 & 5.7 & - & $0.57 \pm 0.03$ & $0.006 \pm 0.003$ & $1.41 \pm 0.04$ \\
CentaurusA & 201.36 & --43.02 & - & $0.1515 \pm 0.0005$ & $0.056 \pm 0.006$ & $0.58 \pm 0.01$ & HESS J0632+057 & 98.24 & 5.81 & - & $0.286 \pm 0.007$ & $0.037 \pm 0.007$ & $1.02 \pm 0.03$ \\
PKS 1424-418 & 216.99 & --42.11 & - & - & - & $0.33 \pm 0.01$ & 1LHAASO J1908+0615u & 287.05 & 6.26 & 0.72 & - & - & $0.49 \pm 0.03$ \\
J0106-4034 & 16.69 & --40.57 & - & - & - & $0.378 \pm 0.007$ & PKS 2145+067 & 327.02 & 6.96 & - & - & - & $0.49 \pm 0.03$ \\
RX J1713.7-3946 & 258.36 & --39.77 & 0.65 & - & - & $0.35 \pm 0.02$ & W 49B & 287.78 & 9.09 & - & $0.236 \pm 0.002$ & $0.042 \pm 0.004$ & $0.88 \pm 0.01$ \\
CTB 37A & 258.56 & --38.52 & - & - & - & $0.40 \pm 0.02$ & OT 081 & 267.89 & 9.65 & - & $0.01 \pm 0.01$ & $0.6 \pm 0.4$ & $0.73 \pm 0.02$ \\
PKS 1454-354 & 224.36 & --35.67 & - & $0.12 \pm 0.04$ & $0.12 \pm 0.02$ & $0.74 \pm 0.07$ & PKS 1502+106 & 226.1 & 10.49 & - & - & - & $0.49 \pm 0.05$ \\
HESS J1741-302 & 265.32 & --30.38 & - & $0.04 \pm 0.04$ & $0.16 \pm 0.02$ & $0.65 \pm 0.03$ & J0242+1101 & 40.62 & 11.02 & - & $1.0 \pm 0.2$ & $0.012 \pm 0.009$ & $2.2 \pm 0.3$ \\
J1924-2914 & 291.21 & --29.24 & - & - & - & $0.50 \pm 0.01$ & 1LHAASO J1959+1129u & 299.82 & 11.49 & - & - & - & $0.49 \pm 0.06$ \\
Galactic center & 266.42 & --29.01 & - & $0.331 \pm 0.004$ & $0.034 \pm 0.003$ & $1.00 \pm 0.01$ & J2232+1143 & 338.15 & 11.73 & - & - & - & $0.49 \pm 0.06$ \\
J2258-2758 & 344.52 & --27.97 & - & - & - & $0.40 \pm 0.03$ & J0121+1149 & 20.42 & 11.83 & - & - & - & $0.52 \pm 0.03$ \\
J1625-2527 & 246.45 & --25.46 & - & - & - & $0.41 \pm 0.03$ & J1230+1223 & 187.71 & 12.39 & - & $0.08 \pm 0.06$ & $0.14 \pm 0.04$ & $0.79 \pm 0.04$ \\
NGC 253 & 11.88 & --25.29 & - & - & - & $0.54 \pm 0.03$ & J0750+1231 & 117.72 & 12.52 & - & $0.21 \pm 0.02$ & $0.110 \pm 0.010$ & $1.12 \pm 0.03$ \\
J0457-2324 & 74.26 & --23.41 & - & - & - & $0.42 \pm 0.02$ & PKS 1413+135 & 214.03 & 13.35 & - & - & - & $0.49 \pm 0.07$ \\
J1833-210A & 278.42 & --21.06 & - & - & - & $0.555 \pm 0.009$ & J0530+1331 & 82.74 & 13.53 & - & - & - & $0.53 \pm 0.03$ \\
J0836-2016 & 129.16 & --20.28 & - & $0.03 \pm 0.03$ & $0.6 \pm 0.4$ & $0.68 \pm 0.05$ & W 51 & 290.75 & 14.14 & 0.12 & - & - & $0.49 \pm 0.07$ \\
J1911-2006 & 287.79 & --20.12 & - & - & - & $0.434 \pm 0.008$ & J2253+1608 & 343.49 & 16.15 & - & - & - & $0.50 \pm 0.04$ \\
J0609-1542 & 92.42 & --15.71 & - & $0.256 \pm 0.009$ & $0.0108 \pm 0.0007$ & $0.84 \pm 0.02$ & PKS 0735+178 & 114.53 & 17.71 & - & - & - & $0.53 \pm 0.03$ \\
SNR G015.4+00.1 & 274.52 & --15.47 & - & - & - & $0.49 \pm 0.01$ & 1LHAASO J1928+1813u & 292.07 & 18.23 & 1.26 & - & - & $1.01 \pm 0.06$ \\
J2158-1501 & 329.53 & --15.02 & - & - & - & $0.43 \pm 0.04$ & J0854+2006 & 133.7 & 20.11 & - & - & - & $0.52 \pm 0.02$ \\
1LHAASO J1825-1337u & 276.45 & --13.63 & - & - & - & $0.50 \pm 0.01$ & RGB J2243+203 & 340.98 & 20.35 & - & $0.24 \pm 0.02$ & $0.05 \pm 0.02$ & $0.91 \pm 0.04$ \\
QSO 1730-130 & 263.3 & --13.1 & - & $0.12 \pm 0.04$ & $0.13 \pm 0.02$ & $0.81 \pm 0.05$ & Crab Nebula & 83.55 & 22.05 & - & $0.29 \pm 0.03$ & $0.063 \pm 0.008$ & $1.16 \pm 0.06$ \\
J1337-1257 & 204.42 & --12.96 & - & - & - & $0.46 \pm 0.03$ & IC 443 & 94.21 & 22.5 & 0.16 & - & - & $0.60 \pm 0.02$ \\
J2246-1206 & 341.58 & --12.11 & - & - & - & $0.45 \pm 0.03$ & PKS 1424+240 & 216.76 & 23.8 & - & - & - & $0.62 \pm 0.06$ \\
PKS 0727-11 & 112.58 & --11.7 & - & $0.29 \pm 0.01$ & $0.054 \pm 0.002$ & $1.01 \pm 0.02$ & MG3 J225517+2409 & 343.82 & 24.19 & - & $1.39 \pm 0.02$ & $0.00012 \pm 0.00004$ & $2.63 \pm 0.03$ \\
TXS 1749-101 & 268.15 & --10.2 & - & - & - & $0.47 \pm 0.02$ & 3HWC J1950+242 & 297.42 & 24.46 & - & - & - & $0.71 \pm 0.01$ \\
HESS J1828-099 & 277.24 & --9.99 & - & $0.07 \pm 0.07$ & $0.5 \pm 0.5$ & $0.71 \pm 0.04$ & 3HWC J1951+266 & 297.9 & 26.61 & 1.7 & - & - & $0.94 \pm 0.05$ \\
J1512-0905 & 228.21 & --9.1 & - & - & - & $0.48 \pm 0.02$ & 1LHAASO J1959+2846u & 299.78 & 28.78 & 0.58 & - & - & $0.65 \pm 0.01$ \\
J0607-0834 & 92 & --8.58 & - & - & - & $0.64 \pm 0.01$ & J0237+2848 & 39.47 & 28.8 & - & - & - & $0.57 \pm 0.04$ \\
QSO 2022-077 & 306.4 & --7.6 & - & $0.11 \pm 0.03$ & $0.12 \pm 0.01$ & $0.76 \pm 0.04$ & J1310+3220 & 197.62 & 32.35 & - & - & - & $0.59 \pm 0.04$ \\
RS Ophiuchi & 267.55 & --6.71 & - & $0.244 \pm 0.007$ & $0.032 \pm 0.003$ & $0.88 \pm 0.01$ & 1LHAASO J2002+3244u & 300.64 & 32.74 & - & - & - & $0.70 \pm 0.01$ \\
J0006-0623 & 1.56 & --6.39 & - & $0.44 \pm 0.01$ & $0.021 \pm 0.001$ & $1.19 \pm 0.02$ & J1613+3412 & 243.42 & 34.21 & - & - & - & $0.60 \pm 0.03$ \\
1LHAASO J1839-0548u & 279.79 & --5.81 & 0.44 & - & - & $0.61 \pm 0.04$ & 1LHAASO J2018+3643u & 304.65 & 36.72 & 0.48 & $0.02 \pm 0.02$ & $0.6 \pm 0.4$ & $0.86 \pm 0.07$ \\
3C279 & 194.05 & --5.79 & - & $0.29 \pm 0.05$ & $0.07 \pm 0.01$ & $1.08 \pm 0.06$ & J2015+3710 & 303.87 & 37.18 & - & - & - & $0.80 \pm 0.07$ \\
J2225-0457 & 336.45 & --4.95 & - & - & - & $0.48 \pm 0.02$ & Mkn 421 & 166.11 & 38.21 & - & $0.450 \pm 0.009$ & $0.043 \pm 0.005$ & $1.43 \pm 0.02$ \\
4FGL J0307.8-0419 & 46.95 & --4.33 & - & - & - & $0.48 \pm 0.02$ & J0927+3902 & 141.76 & 39.04 & - & $0.30 \pm 0.07$ & $0.08 \pm 0.03$ & $1.27 \pm 0.10$ \\
PKS 1741-038 & 266 & --3.83 & - & $0.373 \pm 0.006$ & $0.038 \pm 0.004$ & $1.21 \pm 0.02$ & NGC 4151 & 182.63 & 39.41 & - & - & - & $0.61 \pm 0.02$ \\
1LHAASO J1843-0335u & 280.91 & --3.6 & 0.72 & - & - & $0.48 \pm 0.02$ & Mkn 501 & 253.47 & 39.76 & - & - & - & $0.78 \pm 0.01$ \\
J0339-0146 & 54.88 & --1.78 & - & - & - & $0.54 \pm 0.03$ & J1642+3948 & 250.75 & 39.81 & - & $0.18 \pm 0.04$ & $0.08 \pm 0.01$ & $0.99 \pm 0.02$ \\
HESS J1848-018 & 281.95 & --1.75 & - & - & - & $0.53 \pm 0.04$ & J0555+3948 & 88.88 & 39.81 & - & - & - & $0.90 \pm 0.01$ \\
J0423-0120 & 65.82 & --1.34 & - & - & - & $0.93 \pm 0.06$ & CygOB2 & 307.44 & 41.04 & 2.17 & - & - & $0.64 \pm 0.02$ \\
\hline
\end{tabular}

}
\caption{List of analysed astrophysical objects. Reported are the source's name, equatorial coordinates (RA, $\delta$), best-fit signal strength $\hat{\zeta}$ (i.e. signal flux scale where 1 corresponds to $10^{-8}$ GeV\,cm$^{-2}$\,s$^{-1}$), pre-trial p-value, and 90\% CL upper limits on the flux normalisation factor for a $E^{-2.0}$ spectrum, $\Phi^{90\%}_0$ (in units of $10^{-8}$ GeV\,cm$^{-2}$\,s$^{-1}$).  Dashes (--) in the fitted signal strength and pre-trial p-value indicate sources with a null fitted signal.
\label{t:results}
}
\end{table}

The source with the lowest p-value in this analysis is MG3J225517+2409 with a pre-trial p-value of $(1.2\pm0.4)\cdot10^{-4}$, corresponding to a significance of $(3.69\pm0.09)\sigma$. Its best-fit flux is $E^2\phi=(1.39\pm0.02)\cdot10^{-8}\,\mathrm{GeV}\mathrm{cm}^{-2}\mathrm{s}^{-1}$.

The post probability is estimated using binomial formula $p_\mathrm{post}=1-(1-p_\mathrm{pre})^{N_\mathrm{cand}}$ and it reached $0.013\pm0.004$ or $(2.26\pm0.13)\sigma$ value for MG3J225517+2409. This calculation is compared with the simulation of sky maps with randomly-sampled local directions and times for the same number of events as in data, and repeating the analysis on each new pseudo-sky data. Simulating about $5\cdot10^4$ sky maps, a post-probability p-value of $0.018\pm0.006$, $(2.12\pm0.14)\sigma$) is obtained, which is different from the binomial calculation mostly due to different histogram generation scheme.

The MG3~J225155+2217 candidate appeared in the catalogue following ANTARES analysis of 1255 Fermi 3LAC objects and it reached pre-trial p-value of $1.4\cdot10^{-4}$ using partial dataset~\cite{Albert2021}. In the same analysis, the 3C403 candidate was the most promising among 56 radio galaxies in the field of view of ANTARES reaching a pre-trial p-value of $2.3\cdot10^{-4}$. The candidate catalogue of 106 sources was compiled for the KM3NeT analysis~\cite{ARCA19and21dataset} with a major overlap with the ANTARES catalogue of 169 candidates~\cite{ANTARESdataset}, from which some candidates were removed based on the observed low signal. A simplified estimation of the post-trial is performed to account for the initial selection of the MG3J225517+2409 candidate from 1255 objects and for the following catalogue reduction. The results indicate that the probability of observing a candidate with a pre-trial $p$-value of $1.2 \cdot 10^{-4}$ or lower—as in the case of MG3J225517+2409 is about $0.1$, to be interpreted as a post-trial $p$-value.

The sky map of ANTARES and KM3NeT/ARCA events near MG3 J225517+2409 is shown in Fig.~\ref{fig:Maps}. One of the ARCA21 events has an error box that includes the source location and its signal likelihood is 6.3 times more likely comparing to the one in the background-only scenario.

The KM3NeT/ARCA size and its exposure are growing rapidly, and it will reach the ANTARES performance in the next few years. A joint data analysis is particularly interesting for this period.

\end{document}